\documentclass{aa}
\usepackage{graphics}

\newcommand{\feh}{\mbox{[Fe/H]}}

\newcommand{\beq}{\begin{equation}}
\newcommand{\eeq}{\end{equation}}
\newcommand{\beqa}{\begin{eqnarray}}
\newcommand{\eeqa}{\end{eqnarray}}
\newcommand{\benu}{\begin{enumerate}}
\newcommand{\eenu}{\end{enumerate}}
\newcommand{\bite}{\begin{itemize}}
\newcommand{\eite}{\end{itemize}}
\newcommand{\bdes}{\begin{description}}
\newcommand{\edes}{\end{description}}

\newcommand{\comment}[1]{}

\begin{document}

	\thesaurus{06(10.07.03 omega Centauri, 03.13.8) }

        \title{On the formation and evolution  of the globular cluster
		$\omega$ Centauri}

	\author{Giovanni Carraro$^1$ and Cesario Lia$^2$}
\institute{
$^1$ Department of Astronomy, Vicolo dell'Osservatorio 5, I-35122
        Padua, Italy, ({\tt carraro\char64pd.astro.it}) \\
$^2$ SISSA/ISAS, via  Beirut 2, I-34014, Trieste              
             , Italy ({\tt liac\char64sissa.it})
	}

	\offprints{Giovanni Carraro \\ 
e-mail: carraro@pd.astro.it }

	\date{Received January 5 / Accepted March 14, 2000}

	\maketitle
	\markboth{Carraro and Lia}{$\omega$~Centauri}

\begin{abstract}
By means of N-body/hydrodynamical simulations we model
the evolution of a primordial  $10^{8} M_{\odot}$ 
density peak which ends up in an object 
closely resembling the present
day globular cluster $\omega$~Centauri.\\
We succeed to reproduce the main features of the cluster,
namely the structure, kinematics and  metallicity distribution.\\
We suggest that $\omega$~Centauri might be a cosmological
dwarf elliptical, formed at high redshift, evolved in isolation and 
self-enriched,
and  eventually fallen inside the potential well of the Milky Way,
in agreement with the Searle-Zinn (1978) paradigm for galactic 
globular cluster formation.\\
We finally suggest that $\omega$~Centauri is probably surrounded 
by an extended Dark Matter (DM) halo, for which no observational evidence 
is at present available. We expect that signatures, if any, of the DM halo 
can be found in the kinematics of stars outside about 20 arcmin.


\end{abstract}  

\section{Introduction }
$\omega$~Centauri is one of the most interesting globular cluster in the Milky Way,
and maybe the most studied one (Majewski et al. 1999, Lee et al. 1999, Pancino 1998).
A compilation of its main properties is given in   Table~1.\\
The most striking feature of this cluster is the measured metallicity spread,
which has been interpreted as the evidence of a multiple stellar
population inside it (Norris et. al 1996, Suntzeff \& Kraft 1996).
The precise metallicity distribution function (MDF) 
is nonetheless still disputed. Norris et al (1996) suggest the presence
of a secondary peak at $[Ca/H] \approx -0.9$ roughly 5 times smaller that the main
peak at $[Ca/H]\approx -1.4$. This trend is not confirmed by Suntzeff \& Kraft 
(1996), who claim for a more regular MDF.\\
On the base of a large photometric 
survey Lee et al. (1999) (but see also Ortolani et al 1999) 
analyzed the color distribution of a sample of bright stars,
showing that on the average it has an e-folding trend,
with the presence of several significant
metallicity peaks. However Majewski et al. (1999) arrive at  a somewhat
different conclusion,
showing that the MDF has a gaussian shape with the maximum at $[Fe/H] \approx -1.7$,
and with some evidences of a secondary peak.
Although different, all these analyses point to the common picture of an object
which experienced an irregular self-enrichment over its evolution.\\
Putting together the chemical and kinematical properties Majewski et al. (1999)
claim that  $\omega$~Centauri might be a possible dwarf galaxy relict.\\
In this paper we propose a N-body/gasdynamical model for the formation and evolution
of $\omega$~Centauri, suggesting that this globular cluster can actually be the remnant of 
a dwarf elliptical galaxy, formed  and evolved avoiding strong mergers,
and eventually captured by the Milky Way.\\
To this aim, the plan of this paper is as follows. 
In Sect.~2 we describe our model and the initial
conditions setup; Sect.~3 to 5 are dedicated to the analysis of the structure, 
chemistry and internal
kinematics, respectively,  of our model, and the comparison with $\omega$~Centauri;
in Sect.~6 we investigate about the possible presence of an extended DM halo
around the cluster. Finally Sect.~7 summarizes our results.

\section{Technique and Initial Conditions}
The simulation presented here  has been performed
using the Tree--SPH
code developed by Carraro at al. (1998) and Buonomo et al. (2000).
In this code,  the properties of the gas component are
described by means of the Smoothed Particle
Hydrodynamics (SPH) technique, whereas the
gravitational forces are computed
by means of a hierarchical tree algorithm using
a tolerance parameter $\theta=0.8$ and expanding tree nodes to quadrupole
order. We adopt a Plummer softening parameter.
In SPH each  particle represents a fluid element whose
position, velocity, energy, density etc. are followed in  time and space.
The properties of the fluid are locally estimated by an interpolation
which involves the smoothing length $h_{i}$.
In our code  each particle possesses its own time and space
variable smoothing length $h_{i}$, and evolves with its own time-step.
This renders the code highly adaptive and flexible, and
suited to speed-up the numerical calculations.
Radiative cooling is described by means of numerical tabulations as a
function of temperature and metallicity taken from Sutherland \&
Dopita (1993) and Hollenbach \& McKee (1979). 
This allows us to account for  the effects of variations
in the metallicity among the  fluid elements and for each of these
as a function of time and position.
The chemical enrichment  of the gas-particles caused by Star Formation (SF) and
stellar ejecta is described by means of the  closed-box model applied to
each gas-particle (cf. Carraro et al. 1998 for more details).
SF and Feed-back algorithm are described in great detail by
Buonomo et al. (2000). In brief,  SF
is let
depend on the total mass density - baryonic (gas and stars) and dark matter
- of the system  and  on  the metal--dependent   cooling  efficiency.
Moreover we consider  the effects of energy
(and mass) feed-back from SNe of type II and Ia,
and stellar winds from massive stars.\\

Instead of selecting halos from large cosmological N-body simulations,
we opted to set up a protogalaxy as an 
isolated virialized DM halo, whose density profile is:

\begin{table}
\tabcolsep 0.10truecm
\caption{Basic properties of $\omega$~Centauri. 
Data have been taken from Harris (1996)
compilation but
for the ellipticity, 
taken from Geyer et al (1983).}
\begin{tabular}{cccccc} \hline
\multicolumn{1}{c}{Mass} &
\multicolumn{1}{c}{Core radius} &
\multicolumn{1}{c}{Concentration} &
\multicolumn{1}{c}{$M/L$} &
\multicolumn{1}{c}{$d_{\odot}$} &
\multicolumn{1}{c}{$\epsilon$} \\
 $\times 10^{6}~M_{\odot}$& arcmin & &  & kpc & \\
 2.9 & 2.58 & 1.24 & 3.6 & 5.3 & 0.121\\
\hline
\end{tabular}
\end{table}

\[
\rho(r) \propto \frac{1}{r} .
\]

\noindent
DM particles are distributed inside the sphere according to an
{\it acceptance--rejection} procedure.
Along with positions, we assign also velocities according to:

\[
v(r) \propto  \sqrt (r \times ln (\frac{1}{r})) .
\]

\noindent
which is the solution of the Jeans equation for spherical isotropic
collisionless system for the adopted density profile.

\begin{figure}
\resizebox{\hsize}{!}{\includegraphics{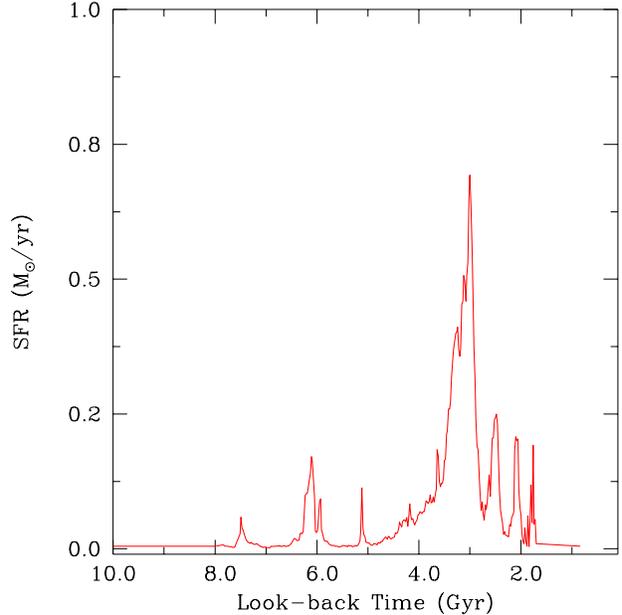}}
\caption{Star Formation Rate as a function of the look-back time.}
\end{figure}

\noindent
The system is let evolve until virial equilibrium is reached.
Gas particles are then distributed homogeneously inside the DM halo with
zero velocity field, thus mimicking the cosmological gas infall inside the
DM potential well (White \& Rees 1978).\\
For the simulation presented here we consider a $9 \times 10^{7} M_{\odot}$
DM halo, sampled by 10,000 particles,
homogeneously filled with $1 \times 10^{7} M_{\odot}$ baryonic mass
in form of metal poor ($Z = 10^{-4}$) gas, sampled by 10,000
particles. This way each gas particle
has a mass of $10^{3} M_{\odot}$.

\section{ The structure}
The evolution of our model is shown in Fig.~1, where we plot the SF as a function of time.
Starting from a $10^{4}$ $^{o}$K  metal poor gas, cooling drives 
baryons towards the center of the 
DM potential well. Part of this gas (about 20\%) is transformed into stars
with a very irregular SF history, whereas the remaining 80\%
is thrown away via a galactic wind mechanism, for which SNe and stellar winds are responsible.\\
The final stars distribution resembles a triaxial object, 
with $a/b=0.859$ and $a/c = 0.967$. 
Correspondingly the ellipticity $(\epsilon = 1 - b/a)$ turns out to be 0.141, not far
from the mean value reported by  Geyer et al. (1983). The stars distribution
is shown in Fig.~2 (onto the X-Z plane) and 3 (onto the X-Y plane). For the sake of an easier
comparison with $\omega$~Centauri the model has been placed at the cluster actual distance
(5.3 kpc).\\
The total mass in stars amount to  about $2 \times 10^{6} M_{\odot}$, sampled
by about 2,000 star particles.
The spherically averaged 
mass density profile derived from our simulation is shown in Fig.~4 . Each filled circle
represents the stars density $\rho_i$ computed as the mass inside a spherical shell $i$
with size the softening $\epsilon$, divided by the shell volume.
For each density estimate $\rho_i$  we plot as error bar
the uncertainty of the density estimate evaluated as the poissonian error related
to the shell population.\\
We perform a fit with a King (1962) profile to check whether our
model recovers the structural properties of $\omega$ Centauri. The cluster
indeed (see Table~1) has a core and tidal radius $r_c$ and $r_t$ equal to
2.6 and 45 arcmin, respectively.\\
By using a King (1962) model (solid overimposed line), 
we found the best fit for $r_c$= 3.2 and $r_t$= 65 arcmin,
which implies a concentration $c = log(\frac{r_t}{r_c})$ = 1.30,
somewhat higher than the value reported in Table~1.
While the value of $r_c$ we obtain agrees with the observed one, the value
of $r_t$ turns out to be much larger than the measured one
for $\omega$ Centauri. In other words we over-estimate the tidal radius.
In our opinion this is due to the fact that in our simplified model
we neglect the effect of the tidal stripping that the cluster
probably did experience when it was accreted by the Milky Way.
This way the outermost stars were probably 
removed reducing the tidal radius.\\
Finally, the central density, computed assuming the $r_c$ derived above,
is about 2660 $M_{\odot}/pc^{3}$. 

\begin{figure}
\resizebox{\hsize}{!}{\includegraphics{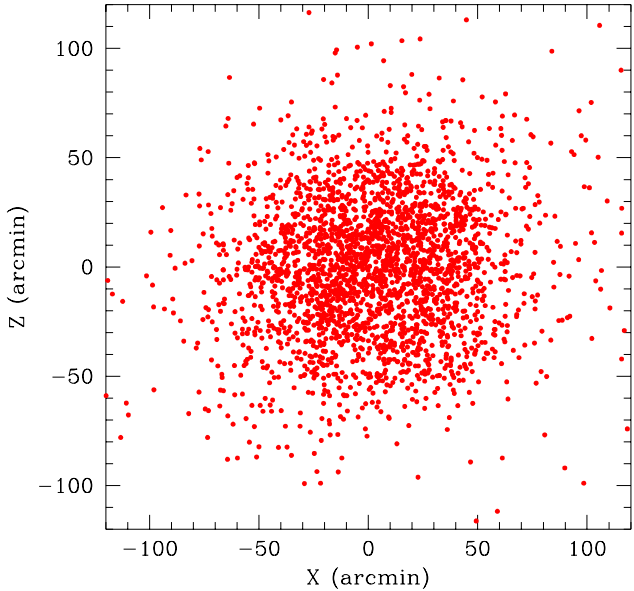}}
\caption{The final distribution of stars in the X-Z plane.
The object has been put at the distance of ~$\omega$ Centauri}
\label{fig_stru}
\end{figure} 

\begin{figure}
\resizebox{\hsize}{!}{\includegraphics{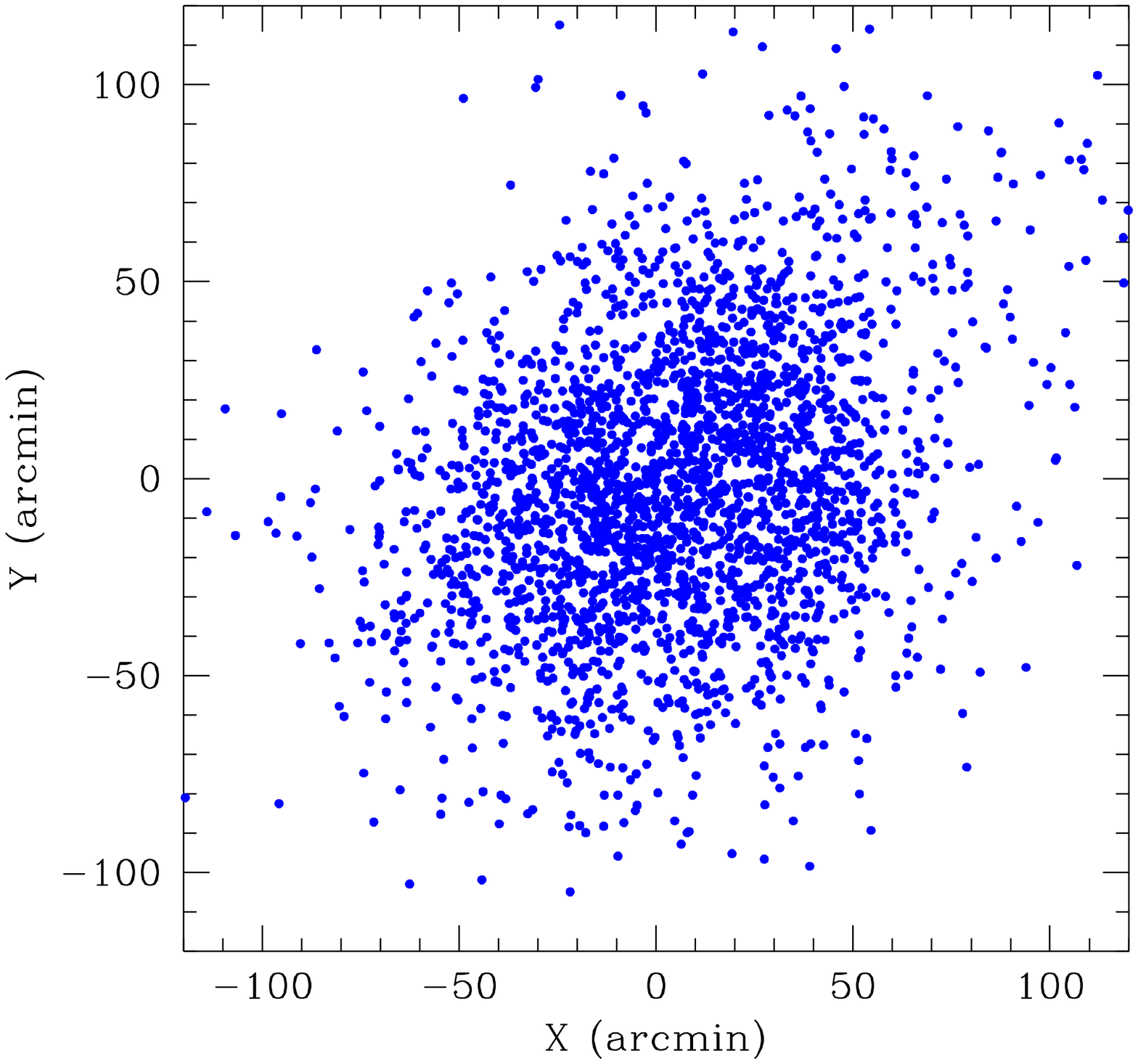}}
\caption{The final distribution of stars in the X-Y plane.
The object has been put at the distance of ~$\omega$ Centauri.}
\label{fig_stru}
\end{figure}

\begin{figure}
\resizebox{\hsize}{!}{\includegraphics{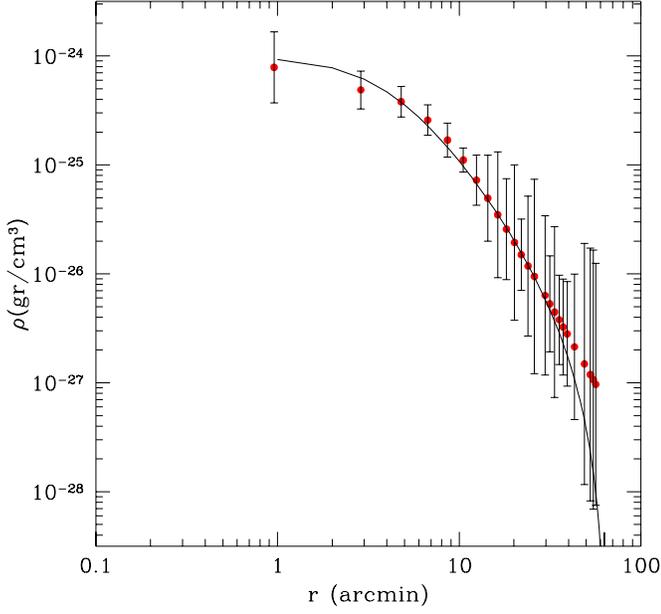}}
\caption{Spherically averaged density profile for our $\omega$~Centauri 
N-body/hydrodynamical model (filled circles).
Overimposed (solid line) is a King (1962) profile for $r_c=3.2$ arcmin and $r_t = 65$ arcmin.
Uncertainties in the density estimates are shown as error bars.}
\end{figure}

\section{Internal kinematics}
The internal kinematics of 
$\omega$~Centauri has been analyzed in great detail by Merritt et al. (1997)
using the radial velocity catalogue obtained with CORAVEL by Mayor et al. (1997).
From this study
it emerges that the cluster has a peak rotational speed of about 8 km/sec
at ~11 pc from the center. Moreover the cluster has a central velocity dispersion 
$\sigma$ = 17 km/sec, the higher one among globular clusters (see Fig.~8 in Merritt et al 1997).\\
From the analysis of our data we find evidences of smaller
rotational speed (about 4 km/sec at 15 pc), whereas the velocity
dispersion profile $\sigma(r)$ (see Fig.~5) shows a central value of about 13 km/sec, denoting
that our model is somewhat colder than $\omega$~Centauri.

\section{Metallicity distribution}
The stars MDF at the end of the simulation is presented in Fig.~6. The bulk of stars
has a metallicity in the range between $\feh = -1.9$ and $\feh -1.6$. 
A secondary peak is observed at  $\feh = -1.2$, about 4 times smaller than the major peak.\\ 
Finally a significant number of stars
has a relatively high metallicity around $\feh = -0.50$. The global shape resembles
the e-folding SF history shown in Fig.~1, with secondary peaks which mark 
successive bursts of star formation.\\
The good agreement we find with the data presented by Majewski et al. (1999)
confirms their suggestion  that $\omega$~Centauri experienced an irregular self-enrichment
over its evolution and may actually be the core of a larger dwarf
elliptical galaxy.

\begin{figure}
\resizebox{\hsize}{!}{\includegraphics{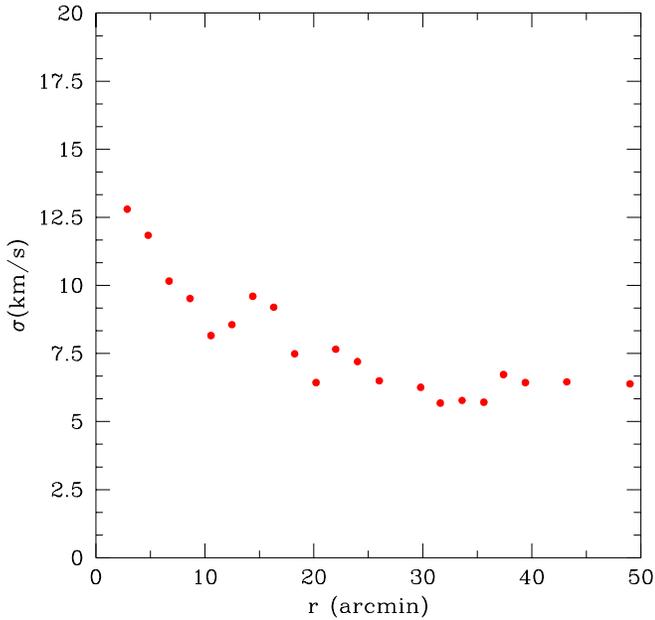}}
\caption{Spherically averaged velocity dispersion profile.}
\label{fig_rho}
\end{figure} 

\begin{figure}
\resizebox{\hsize}{!}{\includegraphics{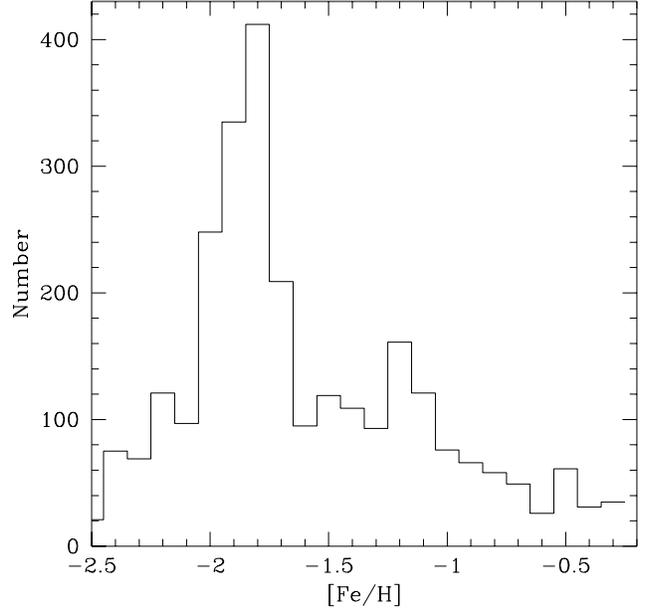}}
\caption{Histogram of the star metallicity distribution at the end of the simulation.}
\label{fig_met}
\end{figure} 

\begin{figure}
\resizebox{\hsize}{!}{\includegraphics{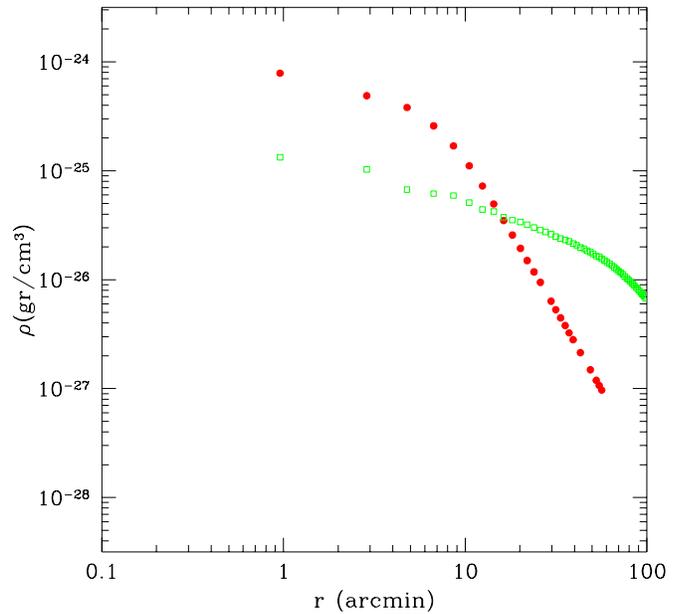}}
\caption{DM  (open squares) and stars (filled circles) density profiles at the end
of the simulation.}
\label{fig_rho}
\end{figure} 

\section{Dark Matter around $\omega$~Centauri?}
Our simulation starts with a virialised DM halo, in whose center  gas
concentrates due to cooling instability,
forming stars. The final stars and DM distribution is shown in Fig.~7,
where open squares refer to DM, and filled circles represent stars.
The inner region of the cluster is dominated by stars up to 20 arcmin, which we call 
the transition radius $r_{tr}$.
In the outer region DM dominates over a distance 6 times larger than the 
transition radius.\\
We expect that the internal kinematics of stars in the inner region is unlikely 
to be influenced by DM. 
On the contrary, stars outside the transition radius $r_{tr}$ (20 arcmin) should show a 
kinematics strongly influenced by DM. A detail spectroscopic study 
of the stars kinematics
in the cluster envelope should be able to confirm or deny
the presence of DM. There is indeed in our model
a trend of the velocity dispersion to weakly 
increase out of the transition radius.

\section{Conclusions}
We have presented a N-body/gasdynamical simulation of the formation and evolution
of the globular cluster $\omega$~Centauri. 
We are able to reproduce the bulk properties of the cluster, namely structure, kinematics
and chemistry assuming that it formed and evolved in isolation, and eventually
fell inside the Milky Way potential well.\\
According to our results and to the dwarf galaxies 
taxonomy proposed by Roukema (1999), $\omega$~Centauri can actually be a cosmological
{\it dwarf by mass}, formed in a high redshift low mass halo, which escaped significant
merging up to the present time. Finally we stress that in order to obtain these results,
an extended DM halo should surround the present day $\omega$~Centauri.

\subsection*{\bf Acknowledgments }
G.C. expresses his gratitude to Prof. 
S. Ortolani for fruitful discussions. The referee,
George Meylan, is acknowledged for important suggestions
which led to the improvement of the paper.
This work has been financed by italian MURST and ASI.

\section*{ References }
\begin{description}
\item Buonomo F., Carraro G., Chiosi C., Lia C., 2000, MNRAS 312, 371
\item Carraro G., Lia C., Chiosi C., 1998, MNRAS 291, 1021
\item Geyer E.H. Hopp U., Nelles B., 1983, A\&A 125, 359
\item Harris W. E. 1996, AJ 112, 1487
\item Hollenbach D., McKee C.F., 1979, ApJS 41, 555
\item King I., 1962, AJ 67, 471
\item Lee Y.-W., Joo J.-M., Sohn Y.-J., Lee H.-C., Walker A.R., 1999,
Nat., in press ({\tt astro-ph/9911137})
\item Majewski S.R., Patterson R.J., Dinescu D.I. et al., 
1999, in "The Galactic Halo: from Globular
Clusters to Field Stars",  35th Liege Astrophysical Colloquium ({\tt astro-ph/9910278})
\item Mayor M., Meylan G., Udry S. et al., 1997, AJ 114, 1087
\item Merritt D., Meylan G., Mayor M., 1997, AJ 114, 1074
\item Ortolani S., Covino S., Carraro G., 1999, in "The Third
Stromlo Simposium: The Galactic Halo", Eds Gibson B.K., Axelrod T.S., 
Putman M.E., ASP Conference Series Vol.~165, 316 
\item Norris  J.E., Freeman K.C., Mighell K.J., 1996, ApJ 462, 251
\item Pancino E., 1998, Master Thesis, Padova University
\item Roukema B.F., 1999, in "XVI Rencontre de Moriond: Dwarf Galaxies and Cosmology", in press
({\tt astro-ph/9806111})
\item Searle L., Zinn R., 1978, ApJ 225, 357
\item Suntzeff N.B., Kraft R.P., 1996, AJ 111, 1913
\item Sutherland R.S., Dopita M.A., 1993, ApJS 88, 253
\item White S.D.M., Rees M., 1978, MNRAS 183, 341
\end{description}

\end{document}